\documentclass{Interspeech2024}
\usepackage{amsmath,graphicx}
\usepackage{amsmath,amsfonts}
\usepackage{algorithmic}
\usepackage{algorithm}
\usepackage{array}
\usepackage{textcomp}
\usepackage{hyperref}
\usepackage{subfig}
\usepackage{stfloats}
\usepackage{url}
\usepackage{verbatim}
\usepackage{tikz}
\usepackage{cite}
\usetikzlibrary{positioning,shapes}

\usepackage{mathtools}
\newcommand{\norm}[1]{\left\lVert#1\right\rVert}
\usepackage{xfrac}

\newcommand{\D}[1]{\mathrm{d}{#1}}

\newcommand{\bmed}{\text{BBED}_{\text{0.08}}}
\newcommand{\blow}{\text{BBED}_{\text{0.01}}}
\newcommand{\bhigh}{\text{BBED}_{\text{0.51}}}

\newcommand{\olow}{\text{OUVE}_{\text{0.01}}}
\newcommand{\omed}{\text{OUVE}_{\text{0.18}}}

\DeclareMathOperator*{\argmin}{arg\,min}
\newcommand{\trsp}{t_{\text{rsp}}}
\usepackage[font=small,skip=0pt]{caption}
\usepackage[nolist, nohyperlinks]{acronym}

\begin{acronym}
\acro{bb}[BB]{Brownian bridge}
\acro{sgm}[SGM]{score-based generative model}
\acro{snr}[SNR]{signal-to-noise ratio}
\acro{gan}[GAN]{generative adversarial network}
\acro{vae}[VAE]{variational autoencoder}
\acro{ddpm}[DDPM]{denoising diffusion probabilistic model}
\acro{stft}[STFT]{short-time Fourier transform}
\acro{istft}[iSTFT]{inverse short-time Fourier transform}
\acro{sde}[SDE]{stochastic differential equation}
\acro{ode}[ODE]{ordinary differential equation}
\acro{ou}[OU]{Ornstein-Uhlenbeck}
\acro{ve}[VE]{Variance Exploding}
\acro{dnn}[DNN]{deep neural network}
\acro{pesq}[PESQ]{Perceptual Evaluation of Speech Quality}
\acro{se}[SE]{speech enhancement}
\acro{tf}[T-F]{time-frequency}
\acro{elbo}[ELBO]{evidence lower bound}
\acro{WPE}{weighted prediction error}
\acro{PSD}{power spectral density}
\acro{RIR}{room impulse response}
\acro{SNR}{signal-to-noise ratio}
\acro{LSTM}{long short-term memory}
\acro{POLQA}{Perceptual Objectve Listening Quality Analysis}
\acro{SDR}{signal-to-distortion ratio}
\acro{ESTOI}{Extended Short-Term Objective Intelligibility}
\acro{ELR}{early-to-late reverberation ratio}
\acro{TCN}{temporal convolutional network}
\acro{DRR}{direct-to-reverberant ratio}
\acro{nfe}[NFE]{number of function evaluations}
\acro{rtf}[RTF]{real-time factor}
\end{acronym}

\def\BibTeX{{\rm B\kern-.05em{\sc i\kern-.025em b}\kern-.08em
    T\kern-.1667em\lower.7ex\hbox{E}\kern-.125emX}}

\name{Bunlong}{Lay}
\name{Timo}{Gerkmann}

\address{
  Signal Processing (SP), University of Hamburg, Germany}
\email{{bunlong.lay@uni-hamburg.de,  timo.gerkmann@uni-hamburg.de}}

\keywords{speech enhancement, diffusion models, stochastic differential equations}

\title{An Analysis of the Variance of Diffusion-based Speech Enhancement}
\begin{document}
\maketitle
\begin{abstract}
Diffusion models proved to be powerful models for generative speech enhancement. In recent SGMSE+ approaches, training involves a stochastic differential equation for the diffusion process, adding both Gaussian and environmental noise to the clean speech signal gradually. The speech enhancement performance varies depending on the choice of the stochastic differential equation that controls the evolution of the mean and the variance along the diffusion processes when adding environmental and Gaussian noise. In this work, we highlight that the scale of the variance is a dominant parameter for speech enhancement performance and show that it controls the tradeoff between noise attenuation and speech distortions. More concretely, we show that a larger variance increases the noise attenuation and allows for reducing the computational footprint, as fewer function evaluations for generating the estimate are required\footnote{Audio examples https://uhh.de/sgmse-variance-analysis}.

\end{abstract}

\section{Introduction}
\label{sec:intro}

The goal of speech enhancement (SE) is to retrieve the clean speech signal from a noisy mixture that has been affected by environmental noise ~\cite{hendriks2013dft}. Traditional methods attempt to leverage the statistical relationships between the clean speech signal and the environmental noise~\cite{gerkmann2018book_chapter}. Various machine learning techniques have been suggested, treating SE as a predictive learning task, as seen in \cite{wang2018supervised, luo2019conv}.

Diverging from predictive approaches, which establish a direct mapping from noisy to clean speech, generative approaches focus on learning a prior distribution over clean speech data. Recently, a category of generative models known as \emph{diffusion models} (or \emph{score-based generative models}) has been introduced to the realm of SE \cite{lu2021study, lu2022conditional, welkerinter2022, journal}. The concept involves gradually adding Gaussian noise to the data through a discrete and fixed Markov chain, referred to as the \emph{forward process}, thereby transforming the data into a tractable distribution like a Gaussian distribution. Subsequently, a neural network is trained to reverse this diffusion process in a so-called \emph{reverse process} \cite{ho2020denoising}. As the step size between two discrete Markov chain states approaches zero, the discrete Markov chain transforms into a continuous-time \ac{sde} under mild constraints. The use of \ac{sde} provides greater flexibility and opportunities compared to methods based on discrete Markov chains \cite{song2021sde}. Notably, \ac{sde}s enable the application of general-purpose \ac{sde} solvers for numerically integrating the reverse process, thereby influencing performance and the number of iteration steps. An \ac{sde} can be viewed as a transformation between two specified distributions, with one designated as the initial distribution and the other as the terminating distribution. For the context of score-based generative models for speech enhancement (SGMSE+) \cite{journal}, recently different SDEs \cite{journal, lay202interspeech, vpidm} have been introduced, with the initial distribution being the clean speech data and the terminating distribution being centered around the noisy mixture. Hence, these SDEs can be thought of as a stochastic interpolation between the clean speech signal and the noisy mixture. Within mild constraints, it is possible to identify a reverse \ac{sde} for each forward \ac{sde}, effectively inverting the forward process \cite{anderson1982reverse, haussmann1986time}. This reverse \ac{sde} starts from a noisy mixture and ends at an estimate of the clean speech. %

Various SDEs \cite{journal, lay202interspeech, vpidm} with different variance and mean evolutions have been recently introduced for diffusion models for the task of SE. The resulting performances and characteristics of these SDEs vary, posing an open question regarding whether the primary contributing factor to performance disparities lies in the mean evolution or the variance evolution. For instance, the variance preserving schedule in \cite{vpidm} outperforms the variance exploding schedule from \cite{journal} with fewer reverse steps needed to solve the reverse SDE. Similar results have been found when comparing the SDE in \cite{lay202interspeech} to the SDE in \cite{journal}.

To address this open question, our paper aims to analyze the different SDEs regarding their variance schedules. We show that the scale of the variance schedule is a decisive factor in the performance of enhancement. More precisely, we show that a larger variance scale increases the amount of noise attenuation in the enhanced signal at the cost of the speech component quality. Vice versa, a lower variance scale results in less noise attenuation, but a better speech component quality. Moreover, we also show that a larger variance allows using fewer reverse steps when solving the reverse SDE, hence, reducing the computational cost of generating the enhanced signal from the noisy mixture. Finally, we experimentally show that different SDEs can perform virtually the same in terms of PESQ when their variance scales are properly optimized.

\section{Diffusion models for Speech Enhancement} \label{sec:sde}

The task of SE is to estimate the clean speech signal $\mathbf S$ from a noisy mixture $\mathbf Y = \mathbf S+\mathbf N$, where $\mathbf N$ is environmental noise. All variables in bold are the coefficients of a complex-valued \ac{stft}, e.g. $\mathbf Y \in \mathbb{C}^d$ and $d=KF$ with $K$ number of \ac{stft} frames and $F$ number of frequency bins.
Following \cite{welkerinter2022, journal}, we model the forward process of the diffusion model with an \ac{sde} defined on $0 \leq t < T_{\text{max}}$:
\begin{equation} \label{eq:fsde}
    \D{\mathbf X_t} =
       \mathbf f(\mathbf X_t, \mathbf Y) \D{t}
        + g(t)\D{{\mathbf w}},
\end{equation}
where $\mathbf w$ is the standard Wiener process \cite{kara_and_shreve}, $\mathbf X_t$ is the current process state with initial condition $\mathbf X_0 = \mathbf S$, and $t$ a continuous diffusion time-step variable describing the progress of the process ending at the last diffusion time-step $T_{\text{max}}$. The functions $\mathbf f(\mathbf X_t, \mathbf Y)$ and $g(t)$ are called drift and diffusion coefficient, respectively. The diffusion coefficient $g$ regulates the amount of Gaussian noise that is added to the process, and the drift $\mathbf f$  affects mainly in the case of linear SDEs the mean of $\mathbf X_t$ (see \cite[(6.10)]{kara_and_shreve}). The process state $\mathbf X_t$ follows a Gaussian distribution \cite[Ch. 5]{sarkkaAppliedStochasticDifferential2019}, called the \emph{perturbation kernel}:
\begin{equation}
\label{eq:perturbation-kernel}
    p_{0t}(\mathbf X_t|\mathbf X_0, \mathbf Y) = \mathcal{N}_\mathbb{C}\left(\mathbf X_t; \boldsymbol \mu(t), \sigma(t)^2 \mathbf{I}\right).
\end{equation}
By Anderson \cite{anderson1982reverse}, each forward SDE as in \eqref{eq:fsde} can be associated to a reverse SDE:
\begin{equation}\label{eq:plug-in-reverse-sde}
    \D{\mathbf X_t} =
        \left[
            -\mathbf f(\mathbf X_t, \mathbf Y) + g(t)^2\mathbf  \nabla_{\mathbf X_t} \log p_t(\mathbf X_t|\mathbf Y)
        \right] \D{t}
        + g(t)\D{\bar{\mathbf w}}\,,
\end{equation}
where
$\D{\bar{\mathbf w}}$ is a Wiener process going backwards in time. In particular, the reverse process starts at $t=T$ and ends at $t=0$. Here $T < T_{\text{max}}$ is a parameter that needs to be set for practical reasons, as the last diffusion time-step $T_{\text{max}}$ is only reached in limit.
The \emph{score function} $\nabla_{\mathbf X_t} \log p_t(\mathbf X_t|\mathbf Y)$ is approximated by a neural network called \emph{score model} $s_\theta(\mathbf X_t, \mathbf Y, t)$, which is parameterized by a set of parameters $\theta$.
Assuming that $s_\theta$ is available, we can generate an estimate of the clean speech $\mathbf X_0$ from $\mathbf Y$ by solving the reverse SDE.

Originally in \cite{welkerinter2022}, it was proposed to use a drift term for the SDE resulting in a mean evolution interpolating between clean and noisy signals. Following this line of research, other SDEs \cite{lay202interspeech, vpidm} interpolating between clean and noisy signals were proposed for the task of SE. To analyze the SE performance of trained score models with different underlying SDEs, here we focus on the SDEs proposed by \cite{journal, lay202interspeech}.

\begin{figure}
    \hspace{-0.2cm}
    \includegraphics{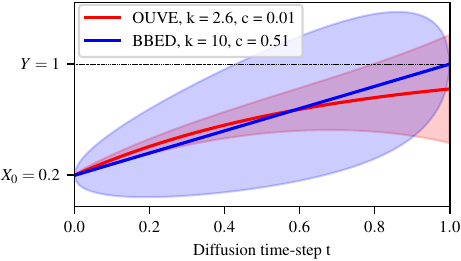}
    \caption{Mean evolutions of BBED and OUVE with parameterization as in \cite{lay202interspeech} and \cite{journal} respectively. Shaded areas indicate the standard deviation of BBED or OUVE.}
    \label{fig:ouve_bbed}
\end{figure}

\subsubsection{Ornstein-Uhlenbeck with Variance Exploding (OUVE)}
In \cite{welkerinter2022, journal} the Ornstein-Uhlenbeck process \cite{kara_and_shreve} was paired with the variance exploding schedule from \cite{song2021sde}. The resulting SDE is called Ornstein-Uhlenbeck with Variance Exploding (OUVE) SDE. The drift coefficient $f(\mathbf X_t, \mathbf Y)$ and diffusion coefficient $g(t)$ for the OUVE SDE are given by
\begin{align} \label{eq:ouve-sde}
    \mathbf f(\mathbf X_t, \mathbf Y) &= \gamma(\mathbf Y-\mathbf X_t), \\ \label{eq:ouve-diffusion}
    g(t) &=  \sqrt{c}k^t,
\end{align}
for $0 \leq t\leq T<T_\text{max} = \infty$ and parameters $\gamma, c,k \in \mathbb R_+$.
The closed-form solution for the mean and variance of the perturbation kernel of the OUVE SDE are given by:

\begin{equation}
    \label{eq:std_ouve}
    \sigma(t)^2 = \frac{
        c\left(k^{2t} - \mathrm e^{-2\gamma t}\right)}
        {2(\gamma+\log(k))}
    \,,
\end{equation}

and 
\begin{equation}
\label{eq:mean_ouve}
   \boldsymbol \mu(t) = \mathrm e^{-\gamma t} \mathbf X_0 + (1-\mathrm e^{-\gamma t}) \mathbf Y
    \,.
\end{equation}

\subsubsection{Brownian Bridge with Exploding Diffusion Term} \label{sec:bbed}
In \cite{lay202interspeech} the Brownian Bridge drift term was paired with an exploding diffusion term and called BBED. The drift coefficient $f(\mathbf X_t, \mathbf Y)$ and diffusion coefficient $g(t)$ for the BBED SDE are given by
\begin{align} \label{eq:bbed-sde}
    \mathbf f(\mathbf X_t, \mathbf Y) &=  \frac{\mathbf Y-\mathbf X_t}{1-t}, \\ \label{eq:bbed-diffusion}
    g(t) &=  \sqrt{c}k^t,
\end{align}
for $0 \leq t < T_\text{max} = 1$ and $k, c \in \mathbb{R}_+$. The mean evolution is given by 
    \begin{equation} \label{eq:ouvp-inter}
    \boldsymbol \mu(t) = (1-t) \mathbf X_0 + t \mathbf Y,
\end{equation}
The variance evolution is given by
\begin{align}\label{eq:bbed:var}
   \hspace{-0.65em} \sigma(t)^2 &=  (1-t)c\left[(k^{2t}-1+t) + \log(k^{2k^2})(1-t)E \right], \\
    E &= \text{Ei}\left[2(t-1)\log(k)\right] - \text{Ei}\left[-2\log(k)\right],
\end{align}
where $\text{Ei}[\cdot]$ denotes the exponential integral function \cite{bender78:AMM}. %

\section{Contribution} \label{sec:contr}
The OUVE and BBED \acp{sde} differ in many aspects which can be observed from Fig. \ref{fig:ouve_bbed}, where we plotted the mean evolutions of OUVE and BBED with shaded areas indicating the standard deviations.
First, the variance evolutions are different, as the variance of OUVE is a strictly increasing function and the variance of BBED first increases and decreases again. This can be seen also in Fig. \ref{fig:ouve_bbed} as the standard deviation of the OUVE SDE (red shaded area) increases and the standard deviation of the BBED SDE (blue shaded area) vanishes at $t=0, T$. Second, the mean evolution of the OUVE SDE is exponential, whereas it is linear for BBED. In \cite{lay202interspeech} the construction of BBED was motivated by reducing the prior mismatch. The prior mismatch is defined as the difference of the \ac{sde}'s mean evolution $\boldsymbol \mu(T)$ to $\mathbf Y$. In \cite{lay202interspeech} it was shown the BBED SDE achieves improved reconstructions over the OUVE SDE and that the BBED SDE has significantly lower prior mismatch than the OUVE SDE. The successful reduction of the prior mismatch by BBED can clearly seen in Fig. \ref{fig:ouve_bbed}, where the red line does not approach $Y$ for $t=T=1$. However, in Section \ref{sec:res} we will show that the dominant factor for the improvement is not the reduction of the prior mismatch, but the amount of Gaussian noise injected during the reverse trajectory as controlled by the variance scale of the SDE. For this, we analyze the variance scale $c$ of the OUVE and BBED SDE and show that with the right choice of the variance scale $c$, the two \acp{sde} perform virtually the same in terms of PESQ in our experiments. Specifically, we reveal the insights:
\begin{enumerate}
    \item When solving the reverse SDE for inference, an SDE parameterized with low variance will not remove as much background noise as an SDE parameterized with a larger variance. In addition, when the variance is too large then it will overaggressively remove energy from the noisy mixture and as a result, the speech component quality of the estimate degrades. This means that the amount of variance of the SDE controls the tradeoff between noise attenuation and speech component quality of the enhanced signal.
    \item An SDE parameterized with a larger variance can reduce the number of network evaluations when solving the reverse SDE for inference. This is because the step size for solving the reverse SDE can be increased, and the reverse starting point $\trsp$ can be reduced to values lower than $T$. Both effects result in fewer steps for solving the reverse SDE and therefore speeding up the inference process.
    \item Even two very different SDEs can lead to similar overall performance for an optimized variance scale.
\end{enumerate}

\section{Experimental setup} \label{sec:exp}
\subsection{Training} \label{sec:exp:train}
For the score model $s_\theta(\mathbf X_t, \mathbf Y, t)$, we employ the Noise Conditional Score Network (NCSN++) architecture (see \cite{journal, song2021sde} for more details). The network is optimized based on denoising score matching:
\begin{equation}\label{eq:training-loss}
      \argmin_\theta \mathbb{E}_{t,(\mathbf X_0,\mathbf Y), \mathbf Z, \mathbf X_t|(\mathbf X_0,\mathbf Y)} \left[
        \norm{\mathbf s_\theta(\mathbf X_t, \mathbf Y, t) + \frac{\mathbf Z}{\sigma(t)}}_2^2
    \right]
\end{equation}
where $\mathbf X_t =  \boldsymbol \mu(t) + \sigma(t) \mathbf Z$ with $\mathbf Z \sim \mathcal N_{\mathbb{C}}(\mathbf 0, \mathbf I)$. We train the network with the ADAM optimizer \cite{kingma2015adam} with a learning rate of $10^{-4}$ and a batch size of 16. For smoothing the network parameters along the training epochs, we employ an exponential moving average of the score model's parameters is tracked with a decay of 0.999 \cite{journal,song2021sde}. We train for 1000 epochs and log the average PESQ value of 7 random files from the validation set during training, selecting the best-performing model for evaluation. %

\subsection{Input representation} \label{sec:exp:data}

Each audio input, sampled at 16 kHz, is converted to a complex-valued \ac{stft}. As in \cite{journal}, we use a window size of 510 samples, a hop length of 128 samples and a periodic Hann window. The input to the score model is cropped randomly to $256$ time frames. A magnitude compression is used to compensate for the typically heavy-tailed distribution of \ac{stft} speech magnitudes~\cite{gerkmann2010empirical}. Each complex coefficient $v$ of the \ac{stft} representation is transformed as $\beta |v|^\alpha \mathrm e^{i \angle(v)}$ with $\beta=0.15$ and $\alpha=0.5$, as in \cite{journal, lay202interspeech}.

\subsection{Datasets}
The publicly available DNS4 dataset \cite{dubey2022icassp} is a large data set designed for SE challenges. This dataset provides clean speech signals and noise clips for mixing. We use only the (anechoic) English clean speech that derives from the LibriVox corpus and mix them with the noise clips with a \ac{SNR} uniformly chosen from $[-5, 10]$ dB. In total, our training set contains 17 hours of paired clean and noisy mixture files, the validation and test set are both 1.7 hours.

\subsection{Stochastic differential equations (SDEs)}
As in \cite{welkerinter2022, journal}, the stiffness parameter $\gamma$, which is only needed for the OUVE SDE in \eqref{eq:ouve-sde}, is set to $1.5$. Both OUVE and BBED depend on the diffusion term parameters $c, k$ in \eqref{eq:ouve-diffusion} and \eqref{eq:bbed-diffusion} respectively. For the OUVE SDE, we set $k=10$ as in \cite{welkerinter2022,journal} and apply a grid-search on the variance scale parameter $c \in [0.01, 0.3]$. Likewise, for the BBED SDE, we fix $k=2.6$ as in \cite{lay202interspeech} and analyze $c \in [0.01, 0.6]$. In addition, we set $T$ to $1.0$ for the OUVE SDE as in \cite{journal} and to $0.999$ for the BBED SDE as proposed in \cite{lay202interspeech}.

\subsection{Sampling} \label{sec:exp:sampling}
\begin{figure}
    \centering
    \includegraphics{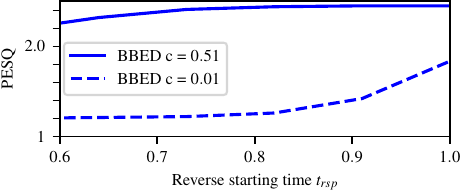}
    \caption{Varying $\trsp$ for trained BBED with fixed reverse step size $\frac{1}{30}$. Demonstrating that higher scaled variances are more robust against increasing the prior mismatch and therefore reducing computational costs.}
    \label{fig:reduce_nfe}
\end{figure}

When solving the reverse SDE, we use the first-order Euler-Maruyama (EuM) method \cite{sarkkaAppliedStochasticDifferential2019}. For this, we denote the reverse start time of the EuM method by $\trsp$, which is the diffusion time from which we start solving the reverse SDE. For the experiments in Fig. \ref{fig:reduce_nfe} we select $\trsp \in \{T, 0.91, 0.82, \dots, 0.19\}$ and use a fixed step size of $\Delta t = 1/30$. 
For the experiments in Tab. \ref{tab:results:dns4} we select $\trsp = T$ and $\Delta t = 1/60$.

\subsection{Metrics} \label{sec:exp:metrics}
We evaluate the performance on the perceptual metric wideband PESQ \cite{rixPerceptualEvaluationSpeech2001}. 
To obtain more insights on the tradeoff between noise attenuation and speech distortions, as in \cite{sigpesq} we evaluate the impact of speech enhancement on clean speech (Speech-PESQ) and noise (noise attenuation (NA)), separately. For this, we first compute a gain function $\mathbf{G} = \mathbf{\hat{S}}/\mathbf{Y}$, where $\mathbf{\hat{S}}$ is the clean speech estimate in the \ac{stft} domain. Then,
we obtain the filtered (time-domain) clean speech $\Tilde{s}$ and noise $\Tilde{n}$ components by applying the gain function $\mathbf{G}$ to the clean speech component $\mathbf{S}$ and the environmental noise component $\mathbf{Y}-\mathbf{S}$, respectively. For Speech-PESQ, we use PESQ applied to $\Tilde{s}$ with the clean speech signal as a reference. For the noise attenuation, we compute the segmental noise-to-filtered-noise ratio in dB (see \cite[Eq. (3)]{sigpesq} for details). 

The use of these filtered metrics \cite{sigpesq, gerkmann2012estimate, lotter} can be motivated as follows. Assume the enhanced signal is of poor speech component quality because it removes energies from the clean speech signal. In this case, the gain function comprises magnitude values close to zero during speech activity, and therefore the filtered speech $\Tilde{s}$ is distorted. Hence, Speech-PESQ yields a poor value. Conversely, a high-quality enhanced signal contains the energy of the clean speech signal, and therefore the gain function has magnitude values close to one during speech activity. Therefore, Speech-PESQ yields a much better value than in the case where speech is distorted by removing energy from the clean speech signal. NA follows the same idea but for the noise signal.

\begin{table}[t]
\resizebox{\columnwidth}{!}{%
\begin{tabular}{l|ccc}
\toprule
Method &  PESQ  & Speech-PESQ & NA [dB] \\
\midrule
\midrule
Mixture & $1.20 \pm 0.21$  &   --- & ---   \\
\midrule

\textbf{} &\multicolumn{3}{l}{Reduced speech component quality, high NA}  \\
BBED, $c=0.51$  & $2.41 \pm 0.73$  &  $2.14 \pm 0.67$  &$\mathbf{33.0 \pm 9.8}$\\
OUVE, $c = 0.73$  & $2.35 \pm 0.65$ & $2.13 \pm 0.65$ & $\mathbf{30.7 \pm 6.5}$ \\
\midrule
\textbf{} &\multicolumn{3}{l}{Good tradeoff between Speech-PESQ and NA}  \\
BBED, $c=0.08$  & $\mathbf{2.43 \pm 0.72}$ & $2.21 \pm 0.69$ &  $29.6 \pm 9.4$  \\
OUVE, $c = 0.18$  & $\mathbf{2.48 \pm 0.73}$ & $2.23 \pm 0.69$ & $29.6 \pm 8.9$  \\
\midrule
\textbf{} &\multicolumn{3}{l}{High speech component quality, reduced NA}  \\
BBED, $c=0.01$  & $2.03 \pm 0.63$ & $\mathbf{2.30 \pm 0.69}$  & $20.9 \pm 8.2$ \\
OUVE, $c = 0.01$ & $2.17 \pm 0.74$ &$ \mathbf{2.39 \pm 0.69}$  &  $20.4 \pm 7.3$\\
\midrule
\bottomrule
\end{tabular} %
}
 \caption{Speech enhancement results (mean $\pm$ std. deviation) obtained for DNS4 based on EuM with N=60 reverse steps.}
\label{tab:results:dns4} 

\end{table}

\begin{figure*}
    \centering
    \includegraphics{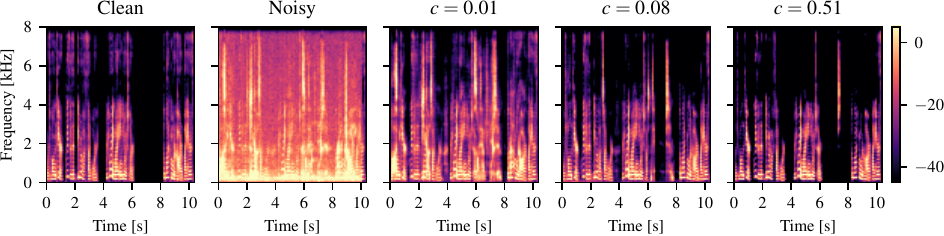}
    \caption{Spectrograms of enhanced signals with BBED with different variance scales $c$ given by models from Tab. \ref{tab:results:dns4}. Spectrograms show the tradeoff between noise attenuation and speech component quality when increasing $c$.}
    \label{fig:spec}
\end{figure*}

\section{Results} \label{sec:res}
In Tab. \ref{tab:results:dns4}, we report enhancement results of the trained score model using either BBED or OUVE for the diffusion process with scales $c$ as discussed in Section \ref{sec:exp}. For brevity, we write $\text{BBED}_{c}$ or $\text{OUVE}_{c}$ when we refer to BBED or OUVE with variance scale $c$.
For generating the enhanced signal with BBED or OUVE in Tab. \ref{tab:results:dns4}, we choose $\trsp = T$ and $N=60$ reverse diffusion steps for the EuM method.

To tackle the first contribution point from Section \ref{sec:contr}, we observe from Tab. \ref{tab:results:dns4} that the speech component quality of the enhanced signal decreases when the variance scale is increased. More precisely, $\blow$ achieves $2.30$ in Speech-PESQ. Increasing the variance scale to $c=0.08$ and $c=0.51$ reduces this value to $2.21$ and $2.14$ in Speech-PESQ, indicating increased speech distortions. In contrast, we observe that the amount of noise attenuation increases when the variance scale increases. More concretely, $\blow$ achieves $20.9$ dB in NA, and BBED with larger variance scale $c=0.08$ and $c=0.51$ increases NA to a much larger value of $29.6$ dB and $33.0$ dB. Likewise, this tradeoff between noise attenuation and speech component quality can also be observed for OUVE in Tab. \ref{tab:results:dns4}. Moreover, in Fig. \ref{fig:spec} we see that the spectrogram produced by $\bhigh$ overaggressively removes energy from the clean speech signal, especially in the high frequencies. In contrast, $\blow$ %
preserves the energy of the clean speech signal but also does not perform as much denoising as $\bhigh$. A reasonable tradeoff between $\blow$ and $\bhigh$ is obtained by $\bmed$.

For the second contribution point from Section \ref{sec:contr}, we argue that an SDE with a larger variance scale is more robust against errors in the reverse trajectory than an SDE with a lower variance scale, as it may help to disguise these errors with a large amount of Gaussian noise. We now show that this is the case when the error source is the discretization of the reverse SDE  or the prior mismatch. First, we discuss how changing the variance scale $c$ affects the performance when the discretization error is increased. To increase the discretization error, we decrease $N$ from 60 to 30 (or equivalently $\Delta t$ increases from $1/60$ to $1/30$) and fix $\trsp = T = 0.999$. Thus, for Fig. \ref{fig:reduce_nfe} we used the same trained models as in Tab. \ref{tab:results:dns4} but with a fixed discretization step size of $\Delta t=1/30$ instead of $\Delta t = 1/60$. From Tab. \ref{tab:results:dns4} we show that $\blow$ with $N=60$ achieves $2.03$ in PESQ. However, in Fig. \ref{fig:reduce_nfe}, we show that $\blow$ with $N=30$ achieves only $1.83$ in PESQ. Therefore, we conclude that when the variance scale $c=0.01$ is relatively low, enhancement performance is sensitive when the discretization error increases. Conversely, we show from 
Fig. \ref{fig:reduce_nfe} that $\bhigh$ with $N=30$ achieves $2.45$ in PESQ which is similar to $2.43$ when $\bhigh$ uses $N=60$ reverse steps in Tab. \ref{tab:results:dns4}.
Hence, we conclude that an SDE with a higher variance scale is less sensitive to discretization errors than an SDE with a lower variance scale. Second, we show that an SDE with a larger variance scale remains more robust when the prior mismatch increases. To increase the prior mismatch, we lower $\trsp$ with fixed $\Delta t = 1/30$. Our experiments in Fig. \ref{fig:reduce_nfe} verify the claim as $\bhigh$ (thick blue curve) remains much more stable than $\blow$ (dotted blue curve) when lowering $\trsp$.

Finally, we observe that the two different SDEs achieve similar performance on the DNS4 dataset when the variance is suitably scaled by comparing $\bmed$ and $\omed$ in all evaluated metrics in Tab. \ref{tab:results:dns4}. Our finding indicates that BBED from \cite{lay202interspeech} outperformed OUVE from \cite{journal} in terms of PESQ because the variance of $\bmed$ in \cite{lay202interspeech} was simply much larger than the variance of $\omed$ (see Fig. \ref{fig:ouve_bbed}). While in \cite{journal} $\olow$ achieved $2.92$ in PESQ on the VoiceBank-Demand \cite{valentini2016investigating}, using the insights of this paper we here report that OUVE with an increased $c=0.08$ achieves $3.11$ in PESQ on the VoiceBank-Demand benchmark. This improvement is due to a higher noise attenuation at the cost of slightly increased speech distortions. Likewise, $\bmed$ achieves $3.08$ in PESQ on the VoiceBank-Demand dataset.

\section{Conclusions} \label{sec:conclusion}
This paper contributes to the understanding of diffusion models for speech enhancement by emphasizing the critical role of the variance scale of the Gaussian noise. Our findings reveal an interesting tradeoff, namely, that a larger variance scale removes more environmental noise at the cost of the speech component quality. Conversely, a smaller variance scale removes less environmental noise but increases in speech component quality. Moreover, we experimentally verified that a larger variance helps to reduce the effect of discretization errors and the prior mismatch. Therefore, when solving the reverse SDE fewer reverse steps are required, hence reducing the computational footprint. Last, we showed that 
with optimized variance scales, OUVE and BBED perform similarly, with OUVE reaching a PESQ of 3.11 on the VoiceBank-Demand benchmark.

\section{Acknowledgements}
The authors gratefully acknowledge the scientific support and HPC resources provided by the Erlangen National High Performance Computing Center (NHR@FAU) of the Friedrich-Alexander-Universität Erlangen-Nürnberg (FAU). The hardware is funded by the German Research Foundation (DFG).

\bibliographystyle{IEEEtran}
\bibliography{ref}

\ifdefined\DeclarePrefChars\DeclarePrefChars{'’-}\else\fi
\begin{thebibliography}{10}
\providecommand{\url}[1]{#1}
\csname url@samestyle\endcsname
\providecommand{\newblock}{\relax}
\providecommand{\bibinfo}[2]{#2}
\providecommand{\BIBentrySTDinterwordspacing}{\spaceskip=0pt\relax}
\providecommand{\BIBentryALTinterwordstretchfactor}{4}
\providecommand{\BIBentryALTinterwordspacing}{\spaceskip=\fontdimen2\font plus
\BIBentryALTinterwordstretchfactor\fontdimen3\font minus \fontdimen4\font\relax}
\providecommand{\BIBforeignlanguage}[2]{{%
\expandafter\ifx\csname l@#1\endcsname\relax
\typeout{** WARNING: IEEEtran.bst: No hyphenation pattern has been}%
\typeout{** loaded for the language `#1'. Using the pattern for}%
\typeout{** the default language instead.}%
\else
\language=\csname l@#1\endcsname
\fi
#2}}
\providecommand{\BIBdecl}{\relax}
\BIBdecl

\bibitem{hendriks2013dft}
R.~C. Hendriks, T.~Gerkmann, and J.~Jensen, \emph{{DFT}-domain based single-microphone noise reduction for speech enhancement: A survey of the state-of-the-art}.\hskip 1em plus 0.5em minus 0.4em\relax Morgan \& Claypool, 2013.

\bibitem{gerkmann2018book_chapter}
T.~Gerkmann and E.~Vincent, ``Spectral masking and filtering,'' in \emph{Audio Source Separation and Speech Enhancement}, E.~Vincent, T.~Virtanen, and S.~Gannot, Eds.\hskip 1em plus 0.5em minus 0.4em\relax John Wiley \& Sons, 2018.

\bibitem{wang2018supervised}
D.~Wang and J.~Chen, ``Supervised speech separation based on deep learning: An overview,'' \emph{IEEE Trans. on Audio, Speech, and Language Proc. (TASLP)}, vol.~26, no.~10, pp. 1702--1726, 2018.

\bibitem{luo2019conv}
Y.~Luo and N.~Mesgarani, ``{Conv-TasNet}: Surpassing ideal time--frequency magnitude masking for speech separation,'' \emph{IEEE Trans. on Audio, Speech, and Language Proc. (TASLP)}, vol.~27, no.~8, pp. 1256--1266, 2019.

\bibitem{lu2021study}
Y.-J. Lu, Y.~Tsao, and S.~Watanabe, ``A study on speech enhancement based on diffusion probabilistic model,'' \emph{IEEE Asia-Pacific Signal and Inf. Proc. Assoc. Annual Summit and Conf. (APSIPA ASC)}, pp. 659--666, 2021.

\bibitem{lu2022conditional}
Y.-J. Lu, Z.-Q. Wang, S.~Watanabe, A.~Richard, C.~Yu, and Y.~Tsao, ``Conditional diffusion probabilistic model for speech enhancement,'' \emph{IEEE Int. Conf. on Acoustics, Speech and Signal Proc. (ICASSP)}, 2022.

\bibitem{welkerinter2022}
S.~Welker, J.~Richter, and T.~Gerkmann, ``Speech enhancement with score-based generative models in the complex {STFT} domain,'' \emph{Interspeech}, 2022.

\bibitem{journal}
J.~Richter, S.~Welker, J.-M. Lemercier, B.~Lay, and T.~Gerkmann, ``Speech enhancement and dereverberation with diffusion-based generative models,'' \emph{IEEE Trans. on Audio, Speech, and Language Proc. (TASLP)}, 2023.

\bibitem{ho2020denoising}
J.~Ho, A.~Jain, and P.~Abbeel, ``Denoising diffusion probabilistic models,'' \emph{Advances in Neural Inf. Proc. Systems (NeurIPS)}, vol.~33, pp. 6840--6851, 2020.

\bibitem{song2021sde}
Y.~Song, J.~Sohl-Dickstein, D.~P. Kingma, A.~Kumar, S.~Ermon, and B.~Poole, ``Score-based generative modeling through stochastic differential equations,'' \emph{Int. Conf. on Learning Representations (ICLR)}, 2021.

\bibitem{lay202interspeech}
B.~Lay, S.~Welker, J.~Richter, and T.~Gerkamnn, ``Reducing the prior mismatch of stochastic differential equations for diffusion-based speech enhancement,'' \emph{Interspeech}, 2023.

\bibitem{vpidm}
Z.~Guo, J.~Du, C.-H. Lee, Y.~Gao, and W.~Zhang, ``Variance-preserving-based interpolation diffusion models for speech enhancement,'' \emph{Interspeech}, 2023.

\bibitem{anderson1982reverse}
B.~D. Anderson, ``Reverse-time diffusion equation models,'' \emph{Stochastic Processes and their Applications}, vol.~12, no.~3, pp. 313--326, 1982.

\bibitem{haussmann1986time}
U.~G. Haussmann and E.~Pardoux, ``Time reversal of diffusions,'' \emph{The Annals of Probability}, pp. 1188--1205, 1986.

\bibitem{kara_and_shreve}
I.~Karatzas and S.~E. Shreve, \emph{Brownian Motion and Stochastic Calculus}, 2nd~ed.\hskip 1em plus 0.5em minus 0.4em\relax Springer, 1996.

\bibitem{sarkkaAppliedStochasticDifferential2019}
S.~Särkkä and A.~Solin, \emph{Applied Stochastic Differential Equations}.\hskip 1em plus 0.5em minus 0.4em\relax {Cambridge University Press}, 2019, no.~10.

\bibitem{bender78:AMM}
C.~M. Bender and S.~A. Orszag, \emph{{Advanced Mathematical Methods for Scientists and Engineers}}.\hskip 1em plus 0.5em minus 0.4em\relax McGraw-Hill, 1978.

\bibitem{kingma2015adam}
D.~P. Kingma and J.~Ba, ``Adam: A method for stochastic optimization,'' \emph{Int. Conf. on Learning Representations (ICLR)}, 2015.

\bibitem{gerkmann2010empirical}
T.~Gerkmann and R.~Martin, ``Empirical distributions of {DFT-domain} speech coefficients based on estimated speech variances,'' \emph{Int. Workshop on Acoustic Echo and Noise Control}, 2010.

\bibitem{dubey2022icassp}
H.~Dubey, V.~Gopal, R.~Cutler, S.~Matusevych, S.~Braun, E.~S. Eskimez, M.~Thakker, T.~Yoshioka, H.~Gamper, and R.~Aichner, ``{ICASSP} 2022 deep noise suppression challenge,'' in \emph{IEEE Int. Conf. on Acoustics, Speech and Signal Proc. (ICASSP)}, 2022.

\bibitem{rixPerceptualEvaluationSpeech2001}
A.~Rix, J.~Beerends, M.~Hollier, and A.~Hekstra, ``Perceptual evaluation of speech quality ({{PESQ}}) - a new method for speech quality assessment of telephone networks and codecs,'' \emph{IEEE Int. Conf. on Acoustics, Speech and Signal Proc. (ICASSP)}, vol.~2, pp. 749--752, 2001.

\bibitem{sigpesq}
S.~Elshamy and T.~Fingscheidt, ``Improvement of speech residuals for speech enhancement,'' \emph{IEEE Workshop on Applications of Signal Proc. to Audio and Acoustics (WASPAA)}, pp. 219--223, 2019.

\bibitem{gerkmann2012estimate}
T.~Gerkmann and R.~C. Hendriks, ``Unbiased mmse-based noise power estimation with low complexity and low tracking delay,'' \emph{IEEE Transactions on Audio, Speech, and Language Processing}, vol.~20, 2012.

\bibitem{lotter}
T.~Lotter and P.~Vary, ``Speech enhancement by map spectral amplitude estimation using a super-gaussian speech model,'' \emph{EURASIP J. Adv. Signal Process}, 2005.

\bibitem{valentini2016investigating}
C.~Valentini-Botinhao, X.~Wang, S.~Takaki, and J.~Yamagishi, ``Investigating {RNN}-based speech enhancement methods for noise-robust text-to-speech,'' \emph{ISCA Speech Synthesis Workshop (SSW)}, pp. 146--152, 2016.

\end{thebibliography}

\end{document}